\documentclass[useAMS,usenatbib]{mn2e}

\usepackage{graphicx,xcolor}
\usepackage{amsmath,amsfonts,amssymb}
\usepackage{aas_macros}

\DeclareMathAlphabet{\mathsc}{OT1}{cmr}{m}{sc}
\def\testbx{bx}%
\DeclareRobustCommand{\ion}[2]{
\relax\ifmmode\ifx\testbx\f@series
{\mathbf{#1\,\mathsc{#2}}}\else
{\mathrm{#1\,\mathsc{#2}}}\fi
\else\textup{\!\!#1\,{\mdseries\textsc{#2}}}\fi}

\title[Filaments and sheets of the WHIM]
{Filaments and sheets of the warm-hot intergalactic medium}

\author[J. S. Klar and J. P. M\"ucket]
{J. S. Klar$^{1}$\thanks{E-mail:jklar@aip.de} 
and J. P. M\"ucket$^{1}$\thanks{E-mail:jpmuecket@aip.de}\\
$^{1}$Leibniz-Institut f\"ur Astrophysik Potsdam (AIP), An der Sternwarte 16, Potsdam, 14482, Germany}

\begin{document}

\date{Accepted 2012 March 6. Received 2012 March 6; in original form 2011 November 16}

\pagerange{\pageref{firstpage}--\pageref{lastpage}} \pubyear{2011}

\maketitle

\label{firstpage}

\begin{abstract}
Filaments, forming in the context of cosmological structure formation, are not only supposed to host the majority of the baryons at low redshifts in the form of the warm-hot intergalactic medium (WHIM), but also to supply forming galaxies at higher redshifts with a substantial amount of cold gas via cold steams. In order to get insight into the hydro- and thermodynamical characteristics of these structures, we performed a series of hydrodynamical simulations. Instead of analyzing extensive simulations of cosmological structure formation, we simulate certain well-defined structures and study the impact of different physical processes as well as of the scale dependencies. In this paper, we continue our work \citep{Klar10}, and extend our simulations into three dimensions. Instead of a pancake structure, we now obtain a configuration consisting of well-defined sheets, filaments, and a gaseous halo. We use a set of simulations, parametrized by the length of the initial perturbation $L$, to obtain detailed information on the state of the gas and its evolution inside the filament. For $L>4$ Mpc, we obtain filaments which are fully confined by an accretion shock. Additionally, they exhibit an isothermal core, which temperature is balanced by radiative cooling and heating due to the UV background. This indicates on a multiphase structure for the medium temperature WHIM. We obtain scaling relations for the main quantities of this core. After its formation, the core becomes shielded against further infall of gas onto the filament, and its mass content decreases with time. In the vicinity of the halo, the filament's core can be attributed to the cold streams found in cosmological hydro-simulations. They are constricted by the outwards moving accretion shock of the halo. Thermal conduction can lead to a complete evaporation of the cold stream for $L > 6$ Mpc/h. This corresponds to halos more massive than $M_\mathrm{halo} = 10^{13} M_\odot$, and implies that star-formation in more massive galaxies can not be supplied by cold streams.
\end{abstract}


\section{Introduction}

At any time during the cosmic history, the majority of baryons is supposed to rest not in stars and galaxies, but in the intergalactic medium (IGM). During the evolution toward low redshifts, this reservoir undergoes a substantial change in properties. At redshifts of $\approx 2$ most of the IGM is photoionized and is at a temperature of about $10^4$ K. This gas can be detected in absorption in the spectra of high redshifted quasars.  At low redshifts, only about 20 \% of the IGM can be in this state \citep{ProchaskaTumlinson08}. The rest can not be accounted for by observational means. The explanation for the absence of these \emph{missing baryons}, which is currently favored in the community, is connected to the formation of the luminous structures in the Universe. While streaming into the gravitational potential wells of the forming structures, the velocity of the gas eventually becomes supersonic. This leads to the formation of shocks, which are able to convert the kinetic energy of the gas into thermal energy. By this \emph{shock heating}, the mean-scale streaming motions produce shock-confined sheets and filaments which contain gas at much higher temperatures compared to the photoionized high redshift IGM. This phase of the IGM is called warm-hot intergalactic medium (WHIM)

The above scenario was developed using hydrodynamic simulations of cosmic structure formation.  Results by \citet{CenOstriker99} suggest that approximately 30 \% to 50 \% of IGM at $z=0$ rest in the 
WHIM. \citet{Dave99,Dave01,Dave10} confirmed these results with simulations of increasing resolution and examined further the interplay between the different phases of the IGM: photoionized, condensed, warm-hot, and hot. Subsequently, more and more simulations were performed, including additional effects like models for galactic superwinds \citep{CenOstriker06} and non-equilibrium effects in the chemical networks \citep{CenFang06}. Using zoom-in simulation techniques \citet{Dolag06} were able to simulate a whole super-cluster filament scenario with high resolution. Further numerical simulations \citep[e.g.,][]{Kang05,Kawahara06,Tornatore10} with different numerical schemes and resolutions also consistently support the described scenario. Recently, the \emph{Over Whelmingly Large Simulations project (OWLS)} \citep{Schaye09}, consisting of a set of large hydrodynamical cosmological simulations, reported predictions for the soft X-ray and UV metal line emission \citep{Bertone10a,Bertone10b} and the absorption by \ion{O}{VI} absorbers \citep{Tepper-Garcia10}.

Due to the high degree of ionization, the observational signature of the WHIM is very weak, in particular with respect to absorption of neutral hydrogen \citep{Cen01,Richter06a,Richter06b}. Therefore, the detection of highly ionized metal lines in the spectra of bright quasars and blazars is much more promising \citep{Hellsten98,Perna98,Fang00,Fang01}. These predictions led to a series of detections \citep{Tripp00,Tripp01,Nicastro02,Fang02,Mathur03,Fujimoto04,Danforth05,Danforth06,Danforth08,Tripp08}, but they are considered to be rather tentative. A detection with sufficiently high signal-to-noise ratio is reported by \citet{Nicastro05a,Nicastro05b}. More recent, \citet{Danforth10} reported the detection of a number of absorbers in the low-redshift IGM in the spectrum toward the BL Lac object 1ES1553+113. In particular, they discuss a triple absorber complex at $z \approx 0.19$, seemingly of multi-phase nature, and correlated with a large-scale filament in the galaxy distribution of the Sloan Digital Sky Survey (SDSS) \citep{Abazajian09}. Additionally, several tentative detections of the WHIM through its metal line X-ray emission are claimed \citep{Kaastra03,Finoguenov03,Nicastro09,Zappacosta10}. 

The WHIM is supposed to reside in shock-confined filaments and sheets of the IGM. These structures are the result of one-and two-dimensional collapse processes. The basic theory for the formation and evolution of structure is already well understood since the sixties of the last century \citep{Doroshkevich64,Zeldovic70,Sunyaev72,Doroshkevich78}. According to these theories, the most probable formation process starts first with the one-dimensional collapse. Only if the underlying dark matter-distribution enters the first caustic, multi-streaming of matter leads to the formation of two-dimensional filaments and eventually knots, which are characterized by matter collapsing in three dimensions. This evolution is closely followed by the gas distribution. 

Recent studies on galaxy formation indicate on another very important aspect of these structures. In the classical picture, galaxies obtain their baryons mainly by smooth accretion of hot gas through their virial shock \citep{White78,Fall80}. Mergers with other halos bring more baryons to the galaxies and influence their internal structure. In high resolution hydrodynamical simulations \citep{Keres05,Ocvirk08,Keres09a,Keres09b,Ceverino09}, however, another aspect of gas accretion is evident. For massive galaxies at redshifts of $z = 2 - 3$ a significant amount of gas streams along the filament directly onto the galactic disk. These \emph{cold streams} seem to penetrate the virial shock and keep a low temperature of $\approx 10^4$ K. This could have significant impact on galaxy formation. The massive supply of cold gas at an early point in galaxy formation does not only influence the star-formation in the galaxy \citep{Dekel09,Bouche10}, but also give an explanation for the unusual morphology of \emph{clump-clusters} and \emph{chain-galaxies} in Hubble Ultra Deep Field \citep{Elmegreen05,Bournaud07,Bournaud09,Agertz09}. It is also discussed in the context of \emph{Lyman-$\alpha$ blobs} at even higher redshifts \citep{Goerdt10,Faucher-Giguere10}.

It is of principal importance to investigate the detailed thermodynamic state and the internal kinematics of the structures which, on the one hand, may hide a large fraction of cosmic baryons at low redshifts, and, on the other hand, may have considerable impact on galaxy formation at higher redshifts. Their detailed modeling, however, is highly demanding for computational astrophysics. The treatment of low-density regions in great detail is difficult. Compared with the numerical handling of high-density matter distributions where adaptive techniques can be applied, for low-density regions, higher overall particle and/or grid number is unavoidable for an appropriate description. In addition, higher resolution calls for a more detailed consideration of local physics, e.g., star formation, feedback, contamination by heavy elements, etc. Altogether the computational effort is immensely more complicated if considered within a cosmological context. Despite of the currently available highly developed computational techniques, an adequate treatment of low-density regions is still at the limit or beyond the near future potential. For these reasons, in order to get informations about filaments and sheets beyond to what simulations of structure formation can tell us, we choose a complementary approach: Instead of analyzing extensive simulations of cosmological structure formation, we simulate certain well-defined structures.

In \citet[hereafter Paper I]{Klar10} we started our investigation with the consideration of the one-dimensional collapse of one perturbation with a given length scale $L$. This scenario is commonly referred to as \emph{cosmic pancake formation} \citep{Zeldovic70,Sunyaev72,Binney77,Bond84,Shapiro85}. We were able to simulate the detailed physics and thermodynamics of the one-dimensional collapse toward a gaseous sheet at extremely high resolution and investigated the impact of radiative cooling, heating by the photoionizing cosmic UV background, non ionization equilibrium (IE), and thermal conduction on the configuration. Above an initial perturbation scale of $\approx 2$ Mpc the collapsing gas shocks. The density and temperature profiles are characterized by the existence of a cold and dense core region which is at thermal equilibrium at a temperature of $T_c\approx 2 \times 10^4$ K. This core is built even before shock formation, and its properties are given by the interplay between radiative cooling and the energetic input from the UV background. The determining variable for the properties of the core is the initial perturbation scale $L$. While non-IE chemistry has no significant effect on the simulation, thermal conduction provides a mechanism to shrink or even evaporate the core.

Based on our one-dimensional simulations in Paper I, we extend our simulation into three dimensions. When considering three perpendicular modes, multi-streaming occurs and leads to the formation of a well-defined filament and a halo. Though the shape of the gas distribution is extremely idealized, it is expected to exhibit the correct density and temperature profiles depending on the initial scale length, again. In addition, in three dimensions it is possible to study the large-scale gas velocities, which themselves are determined by the gravitational potential. The temperature and density profiles for the three-dimensional case are expected to be much more complex even for our highly idealized geometry. The idealized filament-halo scenario is also suited to study the particular hydrodynamical conditions in which accretion via cold streams occurs. In these simulations a relatively cold stream forms along the filament and propagates deep into the halo. The existence of a natural regulation mechanism for the size and existence of a cold core as found in the one-dimensional case might indicate similar restrictions for the cold streams in three dimensions. Compared to cosmological hydro-simulations, our setup, in particular the smaller simulation box and the Jeans-length based refinement strategy, results in a significantly higher resolution in regions of \emph{intermediate overdensity}, and the computational expense is drastically reduced.
\footnote{A modern simulation, i.e. from the OWLS project \citep{Schaye09} which is especially dedicated to the description of the IGM, uses $512^3$ particles in a box of 25 Mpc/h. This results, at an overdensity of $\rho/\bar{\rho} = 100$, in an effective resolution of $\approx 10^6$ particles per (Mpc/h)$^3$. Our simulations use, e.g. 12 refinement levels in a box of 4 Mpc/h. This results, at the same overdensity, in an effective resolution of $\approx 10^9$ cells per (Mpc/h)$^3$, which exceeds the resolution realized by the OWLS simulation by 3 orders of magnitude.}

The paper is organized as follows: In the next section we will recapitulate the theoretical framework for this study, and describe the used code. Since most of these topics were already discussed in Paper I, we would like to refer the reader to the comprehensive description therein. In Sect. \ref{sResults}, we present the result of our three-dimensional simulations. The implications of these results on the physics of gaseous filaments will be further examined in Sect. \ref{sFilament}. In Sect. \ref{sHalo} we use our simulations in order to investigate the different modes of gas accretion on galactic halos. In particular, we analyze the occurrence of cold streams in our simulation. Finally, we discuss our results and possible implications in Sect. \ref{sConclusion}.

\section{Simulations}\label{sTheory}

\subsection{Theoretical Framework}

Since the theoretical framework of this study is described in detail in Paper I, we only review the main concepts. We describe the baryonic component of the Universe as a polytropic gas with an adiabatic exponent of $\gamma = 5/3$. Its evolution is governed by the Euler-equations, which we use in their super-comoving formulation \citep{MartelShapiro98,Doumler09}. We assume a $\Lambda$CDM cosmology close to the WMAP5 results \citep{Komatsu09} with the parameters $\Omega_\Lambda = 0.73$, $\Omega_m = 0.27$, and $H_0 = 71$ Mpc km$^{-1}$ s$^{-1}$. In our simulations we do not compute the collisionless dynamics of the dark matter (DM) self-consistently, but assume it to follow the same distribution as the baryons. This assumption is used for the determination of the gravitational potential only. Since the gravitational potential is relatively insensitive with respect to small deviations of the matter distribution, the dynamics of the gas is almost not affected. 

This is an obviously crude assumption for the description of high density structures like galaxies where the baryon distribution is expected to be  more concentrated than the DM distribution. For lower densities, the clustering of the DM takes place on arbitrary small scales determined by the initial density perturbation spectrum. In contrast, the gas has a finite temperature and pressure. In particular, after the re-ionization the heating by photo-ionization due to the UV background radiation leads to gas temperatures of the order of a few $10^4$ K. The gas pressure causes a much smoother distribution of the gas \citep[see e.g.][]{Harford08}, erasing inhomogeneities on scale lengths smaller than the corresponding Jeans length. However, the locally enhanced DM density, e.g. in small halos, decreases the effective Jeans length. This might lead eventually to the collapse of perturbations in the gas distribution on smaller scales, either initiated or further enhanced by thermal instability. Those small clumps of the gas distribution within the extended filaments cannot be considered within the used approximation. It is a reasonable assumption however, that the small clumps end up finally in hydrodynamical balance with the ambient filament gas. The thermodynamical state of the latter is defined by the UV background radiation and the shock-accreted gas.

Thus, we compute the total matter density by $\rho_\mathrm{tot} = \rho / f_B$, where $f_B = \Omega_b / \Omega_m = 0.16$ denotes the baryon fraction. This is an obviously crude assumption for the description of high density structures like galaxies, but it is sufficiently accurate at low to intermediate densities.

We follow a chemical network consisting of the primordial species \ion{H}{i}, \ion{H}{ii}, \ion{He}{i}, \ion{He}{ii}, \ion{He}{iii}, and electrons in ionization equilibrium. For the computation of the chemical source term, as well as radiative cooling and heating by the cosmic UV background, we include contributions from photoionization, collisional ionization and recombination, dielectric recombination of \ion{He}{ii}, collisional excitation of \ion{H}{i} and \ion{He}{ii}, and bremsstrahlung (three-body processes are neglected). The corresponding rates are taken from \citet{Black81} and \citet{Katz96} and are summarized in Table A.1 of Paper I. We employ primordial mass fractions of hydrogen $\chi_\ion{H}{} = 0.76$ and helium $\chi_\ion{He}{} = 0.24$.

\subsection{Code}\label{sCode}

The simulations of this paper are carried out using the publicly available simulation code \texttt{RAMSES}, introduced in \citet{Teyssier02}. \texttt{RAMSES} is one of the most popular codes in numerical cosmology. It combines an Adaptive Mesh Refinement structure with an N-Body solver for the DM and a hydrodynamic solver for the baryons. The main application of \texttt{RAMSES} lies in the simulation of large-scale stuctures and galaxy formation \citep{Ocvirk08,Agertz09,Agertz10,Goerdt10,Teyssier10,Hahn10}. It is, however, also applied in simulations of star-formation and inter-stellar medium physics \citep{Fromang06,Hennebelle08a,Hennebelle08b}. \texttt{RAMSES} is an extensive code, which can be run in many different configurations. We adapt the code in a way that it resembles \texttt{evora} (the code we developed and used in Paper I) as close as possible. Doing so, we are able to compare our findings from one- and three-dimensional simulations. In the following we discuss our particular choices and discuss differences to the algorithms of \texttt{evora}:
\begin{enumerate}
	\item Similar to \texttt{evora} we use the MUSCL scheme together with the MINMOD slope-limiter and the HLLC Riemann-solver. To overcome the high-Mach-number problem, \texttt{RAMSES} does not use the mechanism described in Sect. B.2 of Paper I. Instead it uses a hybrid conservative/primitive scheme called \emph{pressure fix}. In highly divergent fluxes it computes the internal energy from the primitive variables.
	\item The main advantage of \texttt{RAMSES} (in the context of this work) is its ability to adaptively refine the resolution in certain regions. We use two complementary refinement strategies. The first criteria is the mass inside a cell. A cell is refined, if the mass inside this cell exceeds eight times the mass contained in one cell on the coarsest grid at background density. This criteria ensures a higher resolution in denser regions. However, it is not able to ensure a adequate resolution in regions of intermediate density and lower temperature. Therefore, as a second criteria, a cell is refined if the local Jeans length becomes smaller than ten times the size of the cell.
	\item Due to it's adaptive nature, \texttt{RAMSES} uses a Multigrid Poisson solver to compute the gravitational potential \citep{Press92,Kravtsov97}. The stopping criteria parameter of the Poisson solver is set to $\varepsilon = 10^{-6}$.
	\item The public version of \texttt{RAMSES} offers some very sophisticated models of cooling and heating, star-formation, and feedback processes \citep{Dubois08}, which are too specific and have to many open parameters to be used in the context of this study. In order to ensure comparability to the \texttt{evora} simulations and to keep the full control over the action of the considered physical processes, the cooling-module of \texttt{RAMSES} was replaced by the same cooling model we use in \texttt{evora}.
	\item We use the same heuristic constraints on the different time steps as in Paper I (see Appendix B therein): $C_\mathrm{cfl}=0.5$ for the CFL-condition and $C_a=0.01$ for the constraint limiting the cosmological expansion in one time step. The chemical subcycling uses also $C_\mathrm{cool} = 0.1$ for the computation of the local chemical evolution.
	\item In addition to the described changes, a number of smaller adjustments have been made in order to use the code with the particular physics and initial conditions of this study.
\end{enumerate}

\begin{figure*}
\centering
\includegraphics[width=16cm]{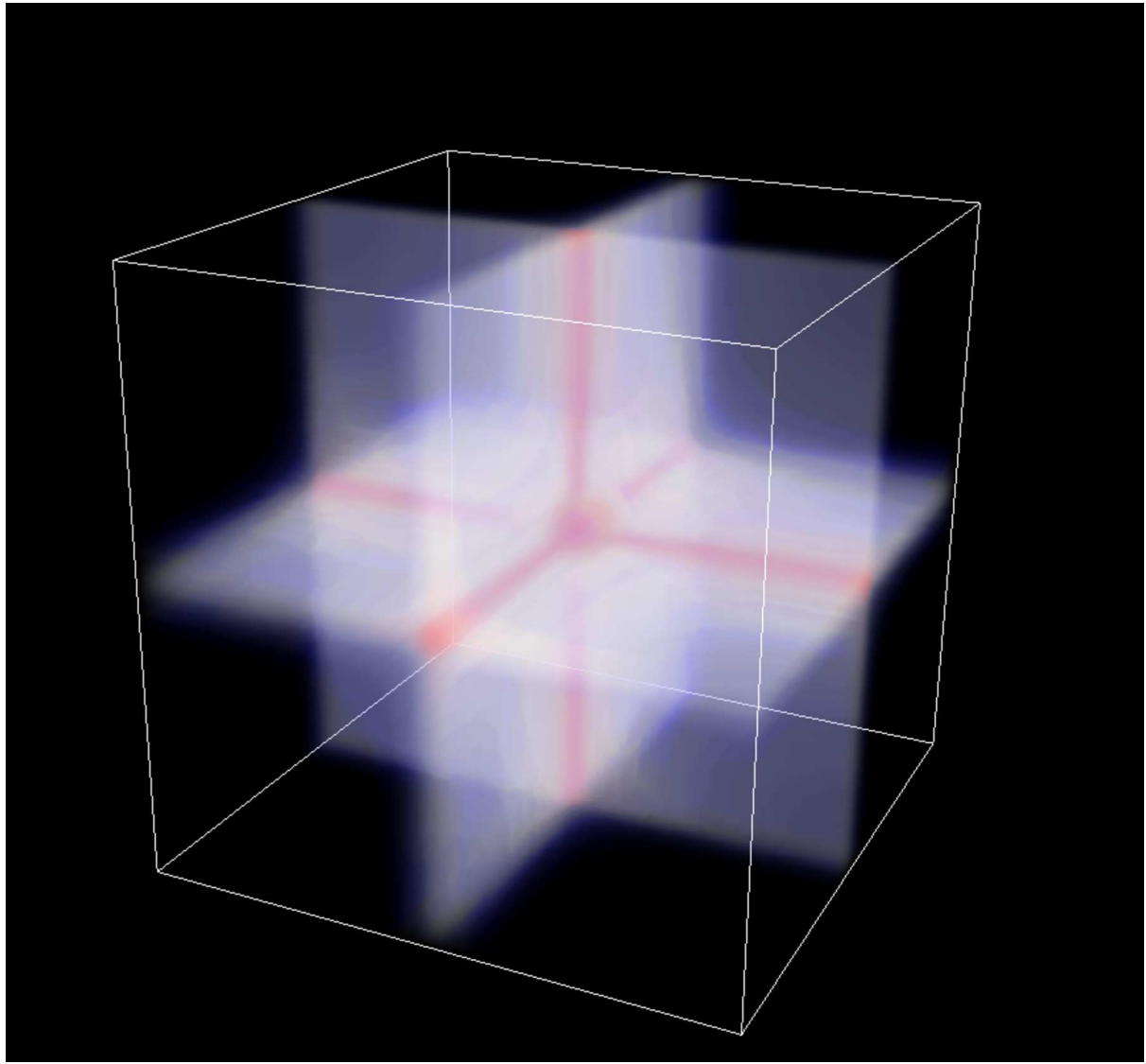}
\caption{Rendering of the density field at $z = 0$ for a three-dimensional simulation using $L = 4$ Mpc. Overdensities in red, underdensities in blue. [\emph{See the electronic edition of the Journal for a color version of this figure.}]
}
\label{fBox}
\end{figure*}

\subsection{Jeans length}\label{sJeans}

In their seminal paper, \citet{Truelove97} demonstrated, that in grid based simulation codes which include hydrodynamics and self-gravitation perturbations of purely numerical origin can cause \emph{artificial fragmentation}. This process can be avoided, if the spatial resolution of the code is sufficient to resolve the local Jeans length of the gas \citep{Jeans02,Jeans28}:
\begin{equation}\label{eTruelove}
	\lambda_\mathrm{J} = \sqrt{\frac{\pi \, c_s^2}{G \, \rho_\mathrm{total}} }
	= \sqrt{ \frac{\pi \, \gamma \, p \vphantom{c_s^2} }{G \, \rho_\mathrm{total} \, \rho_\mathrm{gas}} } \;,
\end{equation}
where $c_s$ denotes the speed of sound of the gas. The fraction of the local Jeans length and the size of a spatial resolution element, called \emph{Jeans number} $J = \lambda_\mathrm{J} / \Delta x$, has to be larger than one. \citet{Truelove97} suggested a Jeans number of $J = 4$. In a cosmological setting, \citet{Ceverino09} found that a Jeans number $J = 7$ is needed. For our particular setup, one-dimensional test simulations show that a Jeans number of $J = 10$ is appropriate.

Using an AMR code, we can use the Jeans length as refinement criterion, increasing the resolution in cold and dense regions if necessary. This can be done, however, only to a certain level of refinement. In order to be able to prevent artificial fragmentation on the finest resolution level, the standard way is to introduce an artificial pressure floor \citep{Machacek01,Agertz09,Ceverino09}. Solving Eq. (\ref{eTruelove}) for the pressure, leads to an equation for a pressure floor:
\begin{equation}\label{ePfloorJeans}
p_{\mathrm{floor}} = \frac{G}{\pi \gamma} \, \Delta x^2 \rho_\mathrm{total} \rho_\mathrm{gas}
\end{equation}
If, during the simulation, the pressure in a cell becomes lower, the code forces the pressure to this value. Of course, this has consequences on the outcome of the simulation: Small scale features in the density distribution will be leveled out. However, in this study we are not interested in these small structures.

\section{Three-Dimensional Collapse}\label{sResults}

\begin{figure*}
\centering
\includegraphics[width=17cm]{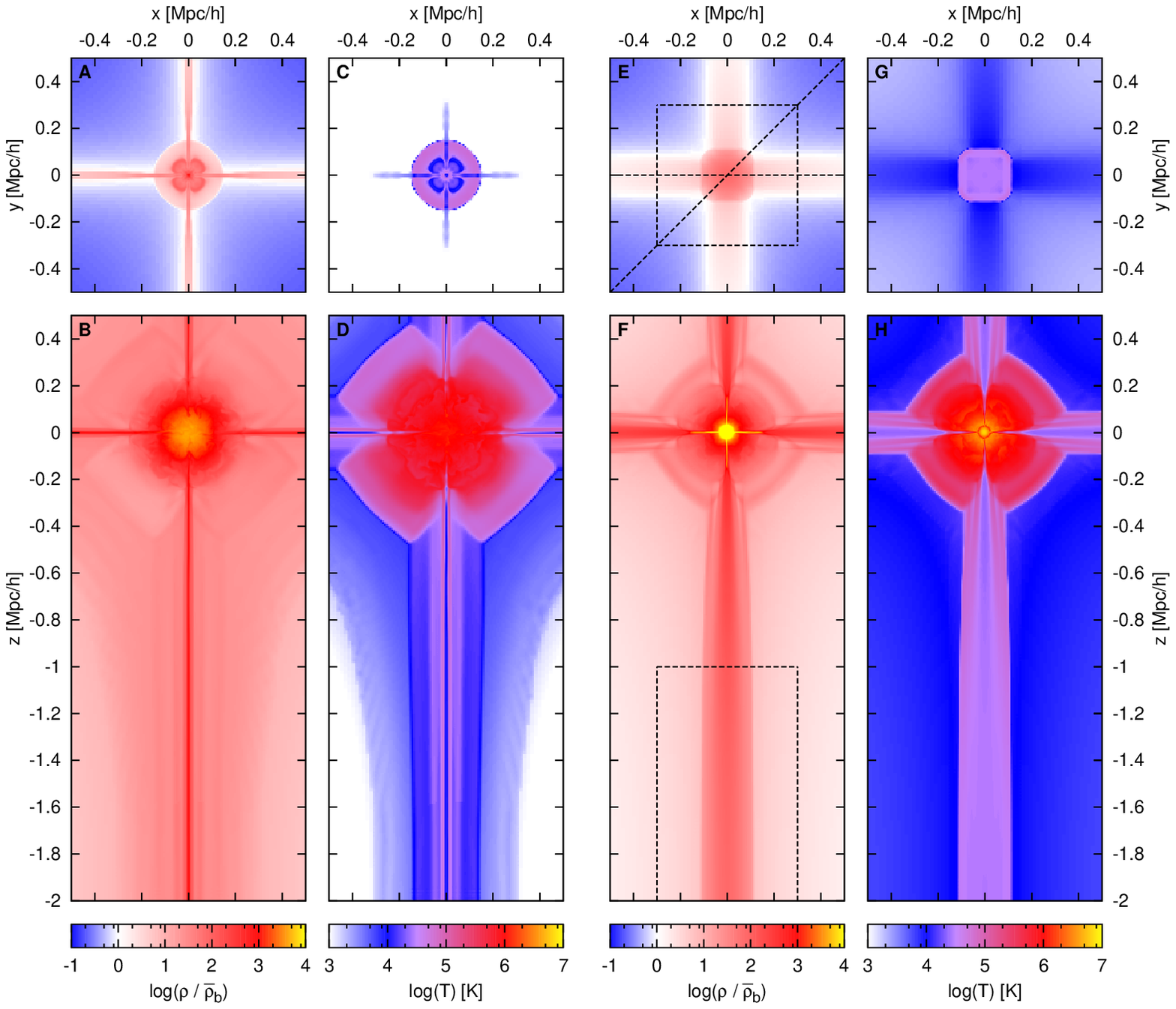}
\caption{Outcome at redshift $z=0$ for a simulation without radiative cooling and heating by the UV background (\emph{Panels} \textbf{A}-\textbf{D}) and with radiative cooling and heating by the UV background (\emph{Panels} \textbf{E}-\textbf{H}). The simulation uses an initial perturbation length of $L=4$ Mpc. \emph{Top Panels:} A slice through the filament of 1 Mpc/h $\times$ 1 Mpc/h the $xy$-plane at the position most distant to the halo; \emph{Bottom panels:} A slice through the halo of 1 Mpc/h $\times$ 2.5 Mpc/h the $xz$-plane; \emph{Panels} \textbf{A},\textbf{B},\textbf{E},\textbf{F}: Density; \emph{Panels} \textbf{C},\textbf{D},\textbf{G},\textbf{H}: Temperature. The color coding of the temperature plot is chosen in way to distinguish photoionized (\emph{blue}), warm-hot (\emph{purple} and \emph{red}) and hot gas (\emph{yellow}). The lines in panel \textbf{E} correspond to the cuts of Fig. \ref{fProfileCool}. The rectangles in panel \textbf{E} and \textbf{F} denote the cuboid volume discussed in Sect. \ref{sStates}. [\emph{See the electronic edition of the Journal for a color version of this figure.}]}
\label{fAdiSlice}\label{fCieSlice}\label{fSlice}
\end{figure*}

\begin{figure*}
\centering
\resizebox{\hsize}{!}{\includegraphics{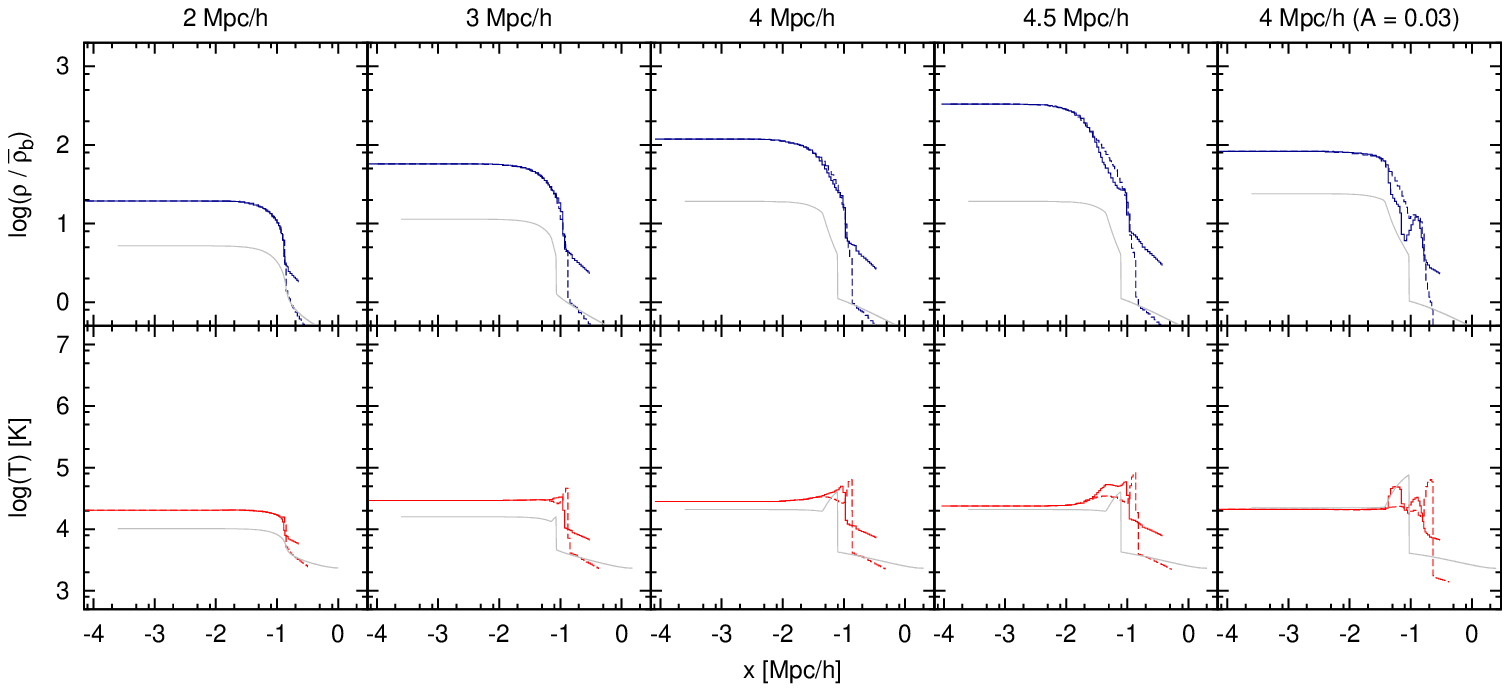}}
\caption{Profiles through the filament for an initial perturbation amplitude of $A= 0.02$ and for a series of perturbation scales of $L=(3,4,4.5,5)$ Mpc/h and for an amplitude of $A= 0.03$ for a perturbation scale of $L = 4$ Mpc/h. Shown are profiles at $z= 0$ including radiative cooling and heating by the uv background, following a line located entirely within the central sheet region (\emph{solid lines}), and following a line coming from the void region and crossing the filament (\emph{dashed lines}). The direction of these lines is shown in panel \textbf{E} of Fig. \ref{fCieSlice} using dashed lines. For comparison the corresponding one-dimensional pancake profiles from Paper I are also shown (\emph{gray lines}). The profiles are displayed using logarithmic coordinate axes.
  \emph{Upper row:} Density profiles; 
  \emph{Lower row:} Temperature profiles.}
\label{fProfileCool}
\end{figure*}

\subsection{Gravitohydrodynamics}

Similar to Paper I, we begin our considerations with simulations \emph{without} the inclusion of radiative cooling and heating by the UV background. We start to run the simulation at redshift of $z=99$. The perturbation on the density field is now decomposed into three sinusoidal perturbations each along the coordinate axes:
\begin{equation}
  \rho = \frac{1}{f_b} \left(
  A_x \cos\big( k_x \, x \big) 
  + A_y \cos\big( k_y \, y \big) 
  + A_z \cos\big( k_z \, z \big) 
  + \bar{\rho} \right)
\end{equation}
The corresponding velocity field is obtained using linear perturbation theory:
\begin{equation}
  \vec{u} = - \frac{f \, \dot{a}}{a} \, \left[
    \frac{A_x \sin( k_x \, x)}{k_x} ,\,
	\frac{A_y \sin( k_y \, y )}{k_y} ,\,
	\frac{A_z \sin( k_z \, z )}{k_z} \right] \;.
\end{equation}
We choose the perturbation scale $L$ to be the same for each wave, and, again as in our one-dimensional consideration, identical to the size of the computational domain in each direction realizing periodic boundary conditions, at once. The choice of that highly symmetrical situation is on purpose by reasons initially discussed. In order to compare with our one-dimensional simulations we choose identical amplitudes along each direction of  $A=A_x=A_y=A_z$, which corresponds to the choice of identical eigenvalues for the initial deformation tensor. The set of initial conditions in the case considered here is unlike the situation as for cosmological simulations. The cosmological simulations generate for each perturbation scale $\propto 1/k$ a set of perturbations at random amplitudes distributed in space realizing a given power spectrum of initial density perturbations $P(k)$. Here we consider the evolution of one particular realization for a single density perturbation on given scale $L$, only. Therefore, this perturbation must be given a well-defined amplitude $A$. This is comparable to the perturbation modes with largest length scales in the cosmological simulations since these modes are also sparsely sampled. However, the value of $A$ affects solely the moment of time of the shock formation. The time of shock formation is closely related to the moment of time when a caustic appears for a one-dimensional collapse of collisionless particles provided the length scale and perturbation amplitude are the same as for the here considered collapse of gas. The relation between the perturbation amplitude and the moment of caustic formation can be obtained from the Zeldovich approximation. At first, we set the amplitude equal to $A=0.02$ in order to obtain a configuration which produces a shock right after the moment $z=1$ (compatible with the WHIM). In order to investigate the possible influence of the value of $A$, we also perform several simulations using $A=0.03$. These configurations produce shocks earlier in cosmic time, and show a more evolved structure at $z=0$. Roughly speaking, the output of these two sets of simulations is nearly identical but delayed by a redshift interval of $\Delta z = 1$. A more detailed analysis will be given below.

The simulations are performed using a coarse grid of $256^3$ cells and up to 4 levels of refinement. A cell is refined, if the mass inside this cell exceeds eight times the mass contained in one cell on the coarsest grid at background density. No Jeans based refinement criteria is applied, yet. We include, however, a pressure floor ensuring the Jeans criteria (for $J=10$) on the finest grid.

As in the one-dimensional case, the collapse starts with an adiabatic contraction of each density perturbation along its initial direction producing a two-dimensional \emph{sheet}. Along the line of intersection of the sheets the densities get superimposed, thus producing an elongated \emph{filament}. The three filaments intersect in one point, leading to a further increase of density resulting in a \emph{halo}. The formation of all these structures is accompanied by shocks if the initial perturbation scales are large enough. Since inside the filaments the density increases faster than in the sheet (also with respect to the one-dimensional collapse), shocks are able to develop much faster there. For the same reason, in the sheet far away from the filament, the collapse is delayed. The density in the halo increases even faster than in the filament, thus the first shock emerges here. Outside these structures (halo, filament, sheet), the space becomes successively devoid of gas and the gas density is lower than the cosmic mean. Thus we will call that region \emph{void}.\footnote{Throughout the further discussion, we will use the nomenclature (void, sheet, filament, halo)  for the above described structures arising in our simulations. We want to emphasize that due to the idealized nature of our simulations, these names should not be directly attributed to the objects, which are usually labeled by these terms in the literature.}. 

For illustration, we show a three-dimensional rendering of the density field for a perturbation length of \mbox{$L=4$ Mpc} in Fig. \ref{fBox}. All the described features, the halo, the filaments, and the sheets can be noticed. In Fig. \ref{fAdiSlice} \textbf{A} - \textbf{D}, we show the density and temperature distribution of the gas in sectional drawings through the simulation box. (Note: the presented sectional drawings are projections with depth of one simulation cell, i.e. we show slices of one cell thickness). The top panels show sections perpendicular to the filament's axis (in $xy$-direction), at the position most distant from the halo, while the lower panels show sectional drawings along the filament (in $xz$-direction) through the halo. The gas distribution can be distinguished by different phases associated with the above described structures:
\begin{itemize}
	\item \emph{Halo:} hot and dense gas confined by strong shocks separating the halo gas from the void region. In addition, some gas at lower temperature is inflowing along the attached filaments. 
	\item \emph{Filament:} warm-hot gas at $10^4-10^5$ K with over-densities of about 10-100
	\item \emph{Sheet:} cold gas $< 10^4$ K and low over-densities $\delta<10$
	\item \emph{Void:} under-dense gas at even lower temperatures
\end{itemize}
Although only gravitohydrodynamics with a polytropic equation-of-state have been assumed so far, streaming features form similar to those which can be noticed in much more elaborated simulations of galaxy formation. In our case, these cold streams are however very narrow and, near the center of the halo, unphysically extended. This happens by the finite resolution and the introduced pressure floor.

\subsection{Cooling and heating}

Following the same schedule as in Paper I, we expand our simulations by the inclusion of radiative cooling and heating by the UV background. Radiative cooling is able to produce cool and dense regions. Therefore, in addition to the mass based refinement we also apply the discussed refinement based on the Jeans-length, now. Otherwise we use the same parameters as in the previous section. 

In Fig. \ref{fCieSlice} \textbf{E} - \textbf{H}, we present section slices with respect to the outcome of such a simulation for $L=4$ Mpc/h similar to the non-radiative case discussed above. Several differences are evident. The heating by the UV background leads to more extended sheets. The filament, forming at their intersection, shows now an extended isothermal core in the center. The filament core has a temperature of $\sim 3 \times 10^4$ K and it exhibits very similar properties compared with the core region forming in the result of the one-dimensional collapse (cp. Paper I). The gas belonging to the cold core streams coherently toward the halo center. During the motion toward the halo, the cross section of the stream, i.e. of the core, gets narrower and narrower. This cold stream is more extended than in the non-radiative case, and penetrates the halo more effectively. 

\subsection{Pressure floor}

As discussed in Sect. \ref{sJeans}, a pressure floor has been introduced to fulfill the Truelove criterion if necessary. This affects only a small region in the very center of the halo and leads to an unphysical high temperature, there. This concerns also the shape and behavior of the cold streams in the vicinity of the halo center. With increasing resolution, the necessary pressure floor is lower. As a consequence in higher resolved simulations, the streams get even narrower toward the halo center. In any case, at these densities our assumption of an optical thin gas will break down eventually, and radiation transport effects would become important. Those could work in both directions, expanding the stream by an additional pressure support since energy can not longer escape from the cold stream gas, or they cold constrict it even more since UV background radiation gets shielded.

A turbulent motion in the shocked region which is visible in all halo plots is a numerical issue. Though in principle, curved shocks can lead to the occurrence of turbulence \citep{Paul11}, our highly symmetric initial conditions are not expected to lead to a final configuration which exhibits this kind of motions. Despite the obtained high refinement level the finite numerical resolution on a rectangular grid together with high density gradients in the vicinity of the halo center seem to excite these non-physical motions. By varying the resolution and the pressure floor, we have proven that neither a significant influence onto the dynamics nor onto the thermal state does exist within the region beyond the innermost cells. Nevertheless, these numerical shortcomings have to be taken into account if considering any results concerning the very central part of the halo. The filament regions distant from the halo are not affected, at all.

\subsection{Mass of the halo}

\begin{figure}
\centering
\resizebox{\hsize}{!}{\includegraphics{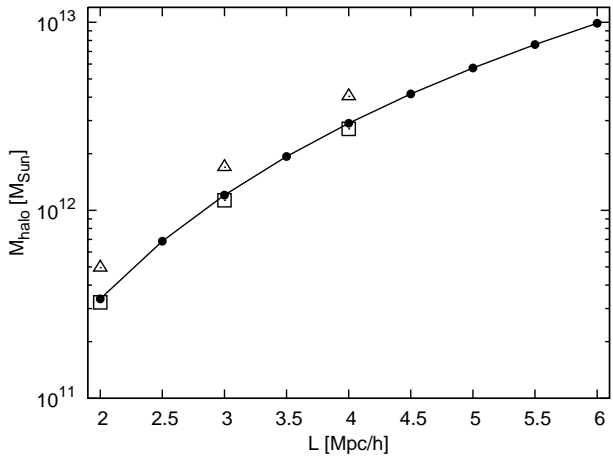}}
\caption{Total halo mass as a function of the initial perturbation length 
for an initial amplitude of $A=0.02$ at $z=0$ (\emph{solid line and circles}),
for $A=0.03$ at $z=1$ (\emph{squares}),
and for $A=0.03$ at $z=0$ (\emph{triangles}). 
For the computation of the mass, a sphere around the halo having a mean density of $\rho/\bar{\rho} = 100$ is considered.}
\label{fMass}
\end{figure}

The amount of matter contained in the void, the sheets, the filaments, and the halo is determined by the initial perturbation scale $L$. In Fig. \ref{fMass} we show the mass of the halo forming in our simulations. Displayed is the total mass of the halo including gas and DM contributions (following our assumptions described above). For an initial amplitude of $A=0.02$ and at a redshift of $z=0$ we obtain masses of $2 \times 10^{11} M_\odot < M_\mathrm{Halo} < 10^{13} M_\odot$. Roughly the same behavior can be found for an higher initial amplitude of $A=0.03$ at an earlier stage of evolution at $z=1$. At $z=0$ the higher amplitude obviously leads to higher halo masses. Since the infall of matter onto the halo is still in the linear regime of cosmological structure formation, the redshift dependence of the halo mass can be well described by the cosmological growth-factor. Since the dependence on redshift and initial amplitude obey almost strong proportionality the corresponding curves appear as to be shifted along the mass axis.

\section{Filament physics}\label{sFilament}

\subsection{Filament profiles}\label{fFilPro}

In this section we will discuss the properties of the filament distant from the halo. In Fig. \ref{fProfileCool} we show profiles of density and temperature along lines perpendicular to the filament and intersecting the filament at the point most distant to the halo position. The profiles given by the solid lines are obtained by following a line in $x$-direction embedded entirely within the central sheet region, while the profiles given by dashed lines are obtained along the diagonal line in $xy$-direction, coming from the void region. These lines are located in the projected slice displayed in the upper panels of Fig. \ref{fSlice} \textbf{E}, and are indicated by the horizontal line for the $x$-direction, and the diagonal line for the $xy$-direction, respectively. Since radiative cooling and heating by the UV background introduces an intrinsic physical scale, we display a set of simulations with different initial perturbation lengths $L=(3,4,4.5,5)$ Mpc/h. The initial perturbation amplitude controls the onset of non-linear evolution and subsequent shock-formation in the simulations. Therefore, we have also included a simulation with a higher initial amplitude of $A = 0.03$ into our considerations. Thus we obtain a more evolved structure at $z=0$.

The profiles show similar properties as the one-dimensional pancakes of Paper I. They are, however, more complex and the reached densities are by an order of magnitude higher. Provided an initial perturbation length of $\approx 4$ Mpc/h a shock is able to form. Then, the obtained profile can be separated into an isothermal core of 10-100 kpc/h and a surrounding shock-heated region. The spatial extension of the core is again determined by the perturbation scale $L$. Larger initial scales $L$ lead to smaller cores. The relation, however, appears to be much steeper compared to the one-dimensional case. This results in a smaller possible scale range for $L$ where a core can exist.

The profiles across the filament obtained for the different directions (one along the  sheet and the other starting from the void region), show no differences for the inner (isothermal) part. For the boundary region of the core however, the density profile along the sheet shows an almost smooth transition from the sheet to the filament region, while for the profile following the direction from the void, a step-like increase can be noticed. In the first case, the corresponding temperature profile exhibits a rather continuous increase toward the filament region, but for the second case, a clear shock is noticeable. 

\subsection{Scaling relations}\label{sScaling}

\begin{figure}
\centering
\resizebox{\hsize}{!}{\includegraphics{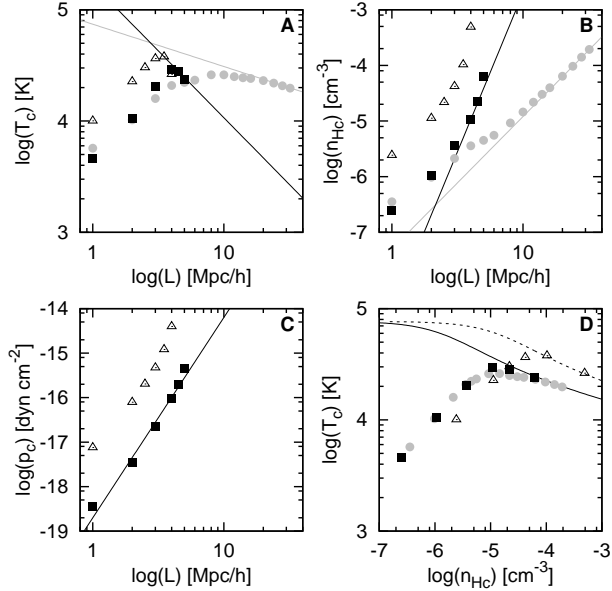}}
\caption{Central temperature $T_\mathrm{c}$ and the central hydrogen number density $n_\mathrm{Hc}$ of the filament for an initial amplitude of $A=0.02$ shown at $z=0$ (\emph{black squares}) and for an initial amplitude of $A=0.03$ shown at $z=1$ (\emph{open triangles}).
\emph{Top Panels:} Dependence on the perturbation length scale $L$
(\emph{Panel \textbf{A}}: Central temperature;
\emph{Panel \textbf{B}}: Central hydrogen number density;
\emph{Panel \textbf{C}}: Central thermal pressure).
The corresponding one-dimensional results from Fig. 6 of Paper I along with the derived one-dimensional scaling relations are given for comparison (\emph{grey points/line}). The three-dimensional scaling relations for the filament, derived in this paper, are indicated by \emph{black lines}.
\emph{Panel \textbf{D}:} Phase space diagram, i.e., the dependence of the central temperature versus the central number density (\emph{black points}) for different perturbation length scales $L$. The solid \emph{black line} denotes the equilibrium temperature $T_\mathrm{eq}$ at redshift $z=0$ and the \emph{dashed line} denotes the equilibrium temperature $T_\mathrm{eq}$ at redshift $z=1$. Again, the corresponding one-dimensional results are given for comparison (\emph{grey points}).}
\label{fScaling}\label{fPhase}
\end{figure}

In Fig. \ref{fScaling} we show the dependencies on the initial perturbation scale $L$ for the temperature $T_\mathrm{c}$ and for the number density of hydrogen $n_\mathrm{Hc}$ in the center of the filament. Shown are results for two sets of simulations: our reference model with an initial amplitude of $A=0.02$, and, for comparison, a simulation with a higher initial amplitude of $A=0.03$. The latter shows faster evolution with time. We therefore compare the output of the simulation employing $A=0.02$ at $z=0$ to the $A=0.03$ simulations at $z=1$. We compare the obtained relations for the three-dimensional case with those obtained for the one-dimensional case as discussed in Sect. 5 of Paper I. Filaments on scales $L < 4$ Mpc/h do not exhibit any shock and their general behavior is similar to what we found for the one-dimensional pancakes on scales less than $2$ Mpc/h. For perturbations evolving over larger scales, shocks appear and significant differences between the three-dimensional and one-dimensional cases can be noticed. In both cases, an increasing scale length $L$ leads to larger central densities, but with a much stronger dependence for the three-dimensional case. The stronger increase in density is accompanied by a stronger decline in temperature. This can be noticed in panel \textbf{A} and \textbf{B} of Fig. \ref{fScaling}. In the three-dimensional case the $T_\mathrm{c}(L)$ and $n_\mathrm{Hc}(L)$ relations are to some extent compressed with respect to the one-dimensional case. Roughly speaking, the one-dimensional scaling relation in the interval $2$ Mpc/h $< L < 24$ Mpc/h corresponds in the three-dimensional case to a much narrower interval $4$ Mpc/h $< L < 5.5$ Mpc/h. A corresponding behavior can be found in the phase-space diagram given in panel \textbf{D} of Fig. \ref{fScaling}. There, we plot the central temperature versus the central hydrogen number density. The behavior of the $T_\mathrm{c}(n_\mathrm{Hc})$ relations differs only marginally for the one- and the three-dimensional cases. Therefore, for the given interval of $L$, we can apply the same arguments for the explanation of the observed scale dependencies as given in \mbox{Paper I}. In particular, the temperature of the core is determined by the balance of radiative cooling and heating by the UV background. We call this equilibrium temperature $T_\mathrm{eq}$. For the simulations employing $A=0.03$ we observe the same behavior at $z=1$. At this point in cosmic time, however, our model for the UV background produces a flux which is by an order of magnitude stronger, resulting in a different equilibrium temperature. Nevertheless, the dependence on the perturbation length remains unchanged. Since the physical densities are plotted the higher values for density at $z=1$ ($A=0.03$) is due to accounting for the different scale factor at higher redshift. For small $L$, as in the one-dimensional case, the reached densities are to low for effective cooling at any time. Thus, after the initial heating by the UV background, the thermal state in the center is determined by the interplay between adiabatic contraction of the gas and the cosmic expansion, only. 

For an one-dimensional collapse, it can be shown that the temperature of the shocked gas scales always as $T\propto L^2$ (cp. with the results of our paper I; see also \citep{Sunyaev72}. In the result, the thermal pressure of the infalling gas, surrounding the filament, scales also $\propto L^2$. This is also true for the thermal pressure inside the sheets, since these are the result of a one-dimensional collapse process. The thermal pressure in the core of the filament, however, is in equilibrium with the the dynamic pressure 
\begin{equation}
	p_\mathrm{dyn,sheet} \propto n_\mathrm{Hc,sheet} \; u_\mathrm{c,sheet}^2
\end{equation}
of the gas compressed into the sheet and streaming onto the filament's core. For the velocity component toward the filament $u_\mathrm{c,sheet}$, one obtains a scaling behavior of $u_\mathrm{c,sheet} \propto L$ from basic hydrodynamical considerations. As shown in Paper I, the scaling behavior of $n_\mathrm{Hc,sheet}$ can be derived from the density dependence of $T_\mathrm{eq}$. For densities of $\rho/\bar{\rho} = 5-15$, appropriate for what we find in the sheets, we obtain $T_\mathrm{eq} \propto n_\mathrm{Hc}^{-0.2}$. As already discussed, the gas pressure outside the sheet scales according to $p_\mathrm{c,sheet} \propto T_\mathrm{eq} \, n_\mathrm{Hc,sheet} \propto L^2$. Following the same line of arguments as in paper I, this leads for the sheets to $n_\mathrm{Hc,sheet} \propto L^{2.5}$.  The state of the core of the filament is determined by the dynamical pressure of the sheet, acting on the core of the filament. It scales therefore like
\begin{equation}
	p_\mathrm{dyn,sheet} \propto L^{4.5} \;.
\end{equation}
since $u_\mathrm{c,sheet} \propto L$.
This relation determines the pressure inside the core of the filament, which therefore shows a much stronger scale dependence as for the core layer of the one-dimensional pancake. This can be noticed in panel \textbf{C} of Fig. \ref{fScaling}. Since the temperature inside the core is determined by the equilibrium temperature, the density is the only variable to adjust a nearly hydrostatic equilibrium. Hence, the relation between $n_\mathrm{Hc}(L)$ for the core of the filament is much steeper than for the one-dimensional pancake. In order to obtain scaling relations for the filament, we can apply the same consideration as before. Since the cores of the filaments are denser than the sheets ($\rho/\bar{\rho} = 100-500$), we have to use a slightly different slope for the equilibrium temperature $T_\mathrm{eq} \propto n_\mathrm{Hc}^{-0.21}$. Using Euler's equation for the estimate of the central pressure $p_\mathrm{c} \propto n_\mathrm{Hc} \phi$, and Poisson's equation $\phi \propto n_\mathrm{Hc} \lambda_\circ^2$, we obtain the relation $p_\mathrm{c} \propto \lambda^2_\circ n_\mathrm{Hc}^2$, where $\lambda_\circ$ is the spatial extend of the core. Using the scaling relation for the pressure inside the filament's core $p_\mathrm{c} \propto L^{4.5}$ derived above, we obtain
\begin{eqnarray}
  n_\mathrm{Hc} &\propto& L^{5.7}  \\
  T_\mathrm{c}  &\propto& L^{-1.2} \\
  \lambda_\circ &\propto& L^{-3.57} \;.
\end{eqnarray}
These analytically derived dependencies approve the outcome of our simulations shown in Fig. \ref{fScaling}. This justifies our assumption about a thermo-dynamical equilibrium state of the filament's core. The relations obtained for the 3-dimensional filament completely differ from the results obtained for the one-dimensional pancake. Though the $n - T$ relation still holds for the equilibrium state of the core where radiative cooling and heating due to the UV-background flux balance each other we get partly much stronger dependencies on the initial perturbations length scale. This is a direct consequence of  the 3-dimensional gas flow structure. In addition, the analytical findings allow for extrapolating to values at larger $L$ where the necessary numerical resolution cannot be reached anymore.

As direct consequence of the much stronger dependence $\lambda_\circ(L)$, a core of reasonable spatial size can only exist for a significant smaller range of perturbation scales. If we use the $L = 4$ Mpc/h case with $\lambda_\circ = 60$ kpc/h as a reference, we obtain a scale of 
\begin{eqnarray}
	L = 14 \, \mathrm{Mpc/h}
\end{eqnarray}
where the core of the filament becomes smaller than 1 kpc/h. Our above made assumption about the tight coupling of baryons and DM might lead to local deviations from the derived scaling relations within the core region. As discussed in Sec. \ref{sTheory} admitting the full initial perturbation spectrum, clustering of the DM on small scales may happen unlimited. These DM clumps may trigger local instabilities of the gas to collapse.These collapsed objects are expected to reach hydrodynamical equilibrium with their ambient medium after some time. The thermodynamical conditions of the ambient medium (core region of the filament) however are determined by the above described processes. Thus, it can be expected that the given scaling relations are valid throughout the filament's core region except for the very local collapsed regions. With other words, averaging over scales larger than the possible gravitational instabilities the derived scaling relations should hold. To follow the evolution of the mentioned instabilities in detail is still beyond the numerical feasibilities.

\subsection{Formation process}\label{sForm}

\begin{figure*}
\centering
\resizebox{\hsize}{!}{\includegraphics{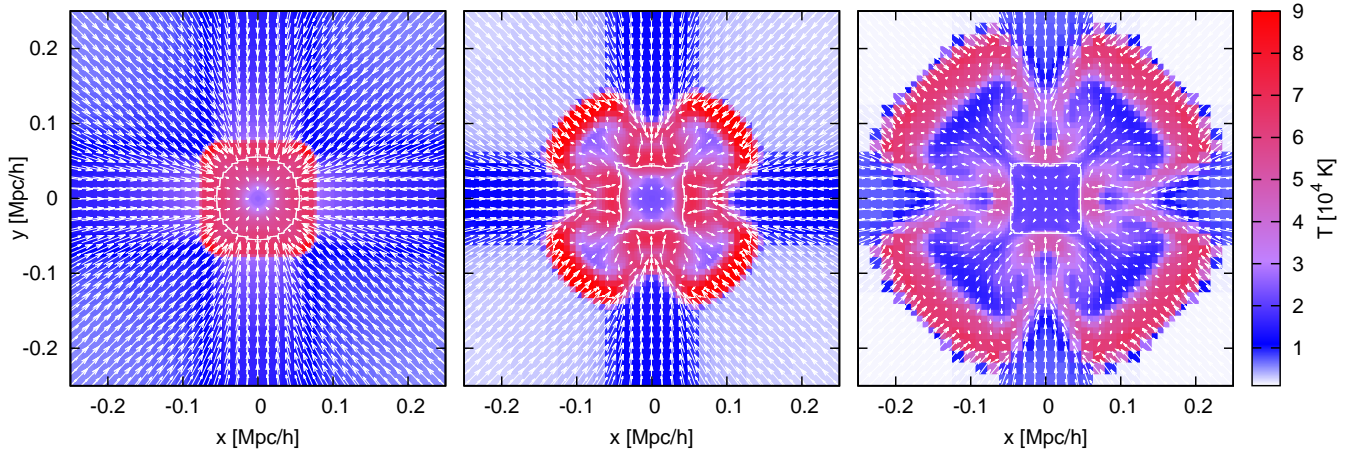}}
\caption{Time evolution of the temperature distribution of the forming filament. Shown are snapshots at expansion factors of a = (0.6,0.8,1.0) corresponding to redshifts of z = (1.0,0.6667,0). The range of the plot is similar to Fig. \ref{fSlice} \textbf{E}. It uses, however, a linear scale. The vectors visualize the gas velocity in the plane of the slice. The white line denotes an over-density of $10^{1.5}$. [\emph{See the electronic edition of the Journal for a color version of this figure.}]}
\label{fTimeSlice}
\end{figure*}

In order to discuss the detailed formation process of the filament we use the simulation employing a higher amplitude of $A = 0.03$. For this initial amplitude, the filament reaches the stage of non-linear evolution and shock-formation before $z=1$. Thus we are able to follow the formation of the filament for a longer time and up to a later stage of evolution. In Fig. \ref{fTimeSlice}, we show the temperature distribution in slices analogously to the upper right panel of Fig. \ref{fCieSlice} at different moments in time. The left-hand panel shows the evolutionary stage at the redshift $z=1$ right after the shock has formed. The subsequent evolution is shown on the middle panel at $z=0.667$, and the right-hand panel at $z=0$.

Before shock formation, matter accretes onto the filament almost equal concentrically but slightly enhanced along the planes given by the sheets where the density is higher. Density and temperature in the filament increase adiabatically, until a shock begins to form. Because of the much lower sound speed in the void region, a shock is forming first at the interface between the forming filament and the void. Further accretion can be distinguished by two phases:
\begin{itemize} 
	\item Material accreting from the void, which is shock-heated to warm-hot temperatures ($T>10^{5}$).
	\item Material accreting along the sheet, which is heated by the transformation of its gravitational energy if the gas is moving toward the center, only. 
\end{itemize}
This corresponds to our findings in Sect. \ref{fFilPro}, where the profile exhibits a shock at the interface to the void region but not in direction to the attached sheet. Noticeable, the velocity field is a good tracer for the shock, declining rapidly behind the discontinuity.

The isothermal-core of the filament is not influenced by this further accretion. The material accreted from the sheet is diverted toward the void resulting in an effective outflow from the filament. This leads to a cold diffuse clover-leaf-shaped region around the filament. A similar behavior is found in the one-dimensional pancakes, where the subsequent accretion also does not influence the central distribution, but rather accumulates gas around the core. The observed outflow however is only possible in two or more dimensions. Summarizing, the gas inside the filament can be distinguished into three-phases: a warm-hot shocked gas layer, an isothermal core, and a cold diffuse region in between. After some moment of time, the core gets "shielded" against further infall of gas and remains thus a limited reservoir of cold gas streaming towards the halo. Besides the described reflection of infalling gas, the hot gas in the immediate vicinity of the cold core is not able to cool anymore due to the large cooling time.

\subsection{Time evolution of the core}\label{sStates}

\begin{figure}
\centering
\resizebox{\hsize}{!}{\includegraphics{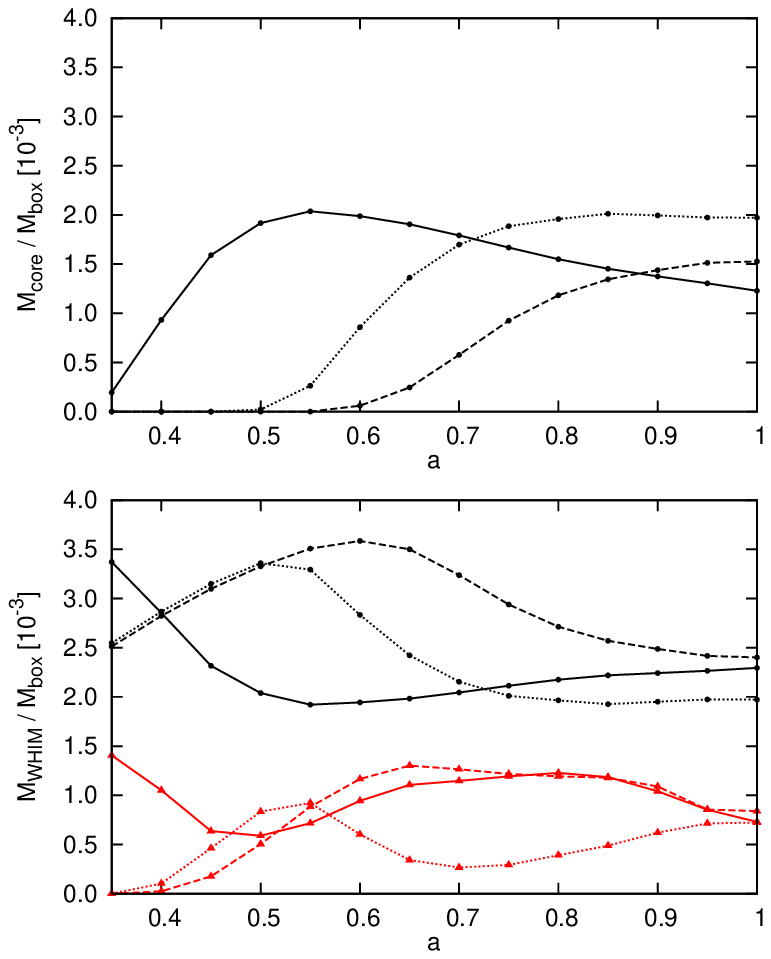}}
\caption{Time evolution of the mass fraction of the gas which resides in the isothermal core and has an over-density of $\rho / \bar{\rho} > 10^{1.5}$ (\emph{upper panel}), and for the gas outside the core having an overdensity of $\rho / \bar{\rho} < 10^{1.5}$ (\emph{lower panel, upper curves}) for the cuboid volume around the filament described in the text. Also displayed is the evolution of the mass fraction of the gas at temperatures $T > 10^{4.5}$ K and $\rho / \bar{\rho} < 10^{1.5}$ (\emph{lower panel, lower curves}), which characterize the evolution of the hot gas envelope around the core. Shown are the curves for the simulations using 
$L = 4$ Mpc (\emph{dashed line}),
$L = 5$ Mpc (\emph{dotted line}),
and $L = 4$ Mpc, but with an initial amplitude of $A = 0.03$, thus forming shock earlier (\emph{solid line}).}
\label{fStatesTime}
\end{figure}

Next, we are considering the time evolution of the amount of gas which resides in the core of the filament. To that purpose, we extract a cubical volume of size 0.6 Mpc/h $\times$ 0.6 Mpc/h $\times$ 1.0 Mpc/h embedding a representative part of the filament region. This cuboid is centered in $x$- and $y$-direction.  In $z$-direction it covers the range from the edge of the box up to half of the distance to the halo center. The rectangular projections of the cuboid are shown in Fig. \ref{fSlice} \textbf{E} and \textbf{F} (dashed lines). For the chosen volume we compute the fractions of the gas at different states. We first construct a phase-space diagram \citep[$T$ versus $\rho/\bar{\rho}$, cp.][]{Valageas02}, weighted by the mass of the individual cells. Then, we separate the gas contributions according to the different phase states. We use a density of $\rho/\bar{\rho} = 10^{1.5}$ as the threshold marker in order to distinguish between the true core gas and the cold and diffuse outflow region. In Fig. \ref{fStatesTime} we display the time evolution for the dense gas forming the isothermal core as well as the less dense envelope around it (computed from subsequent simulation snapshots). The evolution curves are displayed for simulations with $L=(4,5)$ Mpc/h for an initial amplitude of $A=0.02$, as well as for the simulation with $L=4$ Mpc/h at the higher initial amplitude $A=0.03$.

For the simulations employing the lower initial amplitude, the fraction of dense gas, displayed in the upper panel, rises monotonically but with decreasing slope till $a = 1$. For larger $L$ more cold gas accumulates in the core, which is in agreement to our findings in Sect. \ref{sScaling}. The higher initial amplitude of $A=0.03$ allows to follow up the evolution up to later stages. From Fig. \ref{fStatesTime} one can read, that the use of a higher initial amplitude is shifting the plot for $L=4$ Mpc/h to the left along the time axis. The plot shows that the fraction of the cold and dense gas decreases at later moments of time. 

For the fraction of gas in the gaseous envelope, displayed in the lower panel of Fig. \ref{fStatesTime}, we observe an inversed behavior in respect to the mass fraction of the gas inside the core. The part of the envelope, which has a temperature of $T > 10^{4.5}$ and can therefore be considered warm-hot, however, follows roughly the evolution of the core. Deviations are caused by the complicated temperature distribution around the core. When the cold core emerges in the center of the forming filament, the fraction of the warm-hot enveloping gas is decreasing. Later, when the core is formed and is shielded against further infall, the mass fraction of the ambient warm-hot gas is increasing, again. At later stages, a decrease of the warm-hot fraction can be noticed. This is caused by the adiabatic cooling due to the cosmological expansion.

Hence, from a given point in time, the filament evolution leads to a continuous decrease of the mass fraction of the isothermal core. This stage of evolution, however, is only reached untill $z=0$ for filaments forming from perturbations with a high enough initial amplitude. The process can be explained by the process described in Sect. \ref{sForm}. Starting from the denser regions next to the halo, the shock encloses more and more of the filament, until no cold gas can accrete on it from the void. The remaining reservoirs of cold gas are the sheets. However, this gas does not reach the core any more, but it is redistributed around it. Since, with time, the cold gas in the core is drained toward the halo, the amount of cold dense gas in the filament is decreasing.

\section{Accretion onto the halo}\label{sHalo}

\subsection{Formation process}

\begin{figure*}
\centering
\resizebox{\hsize}{!}{\includegraphics{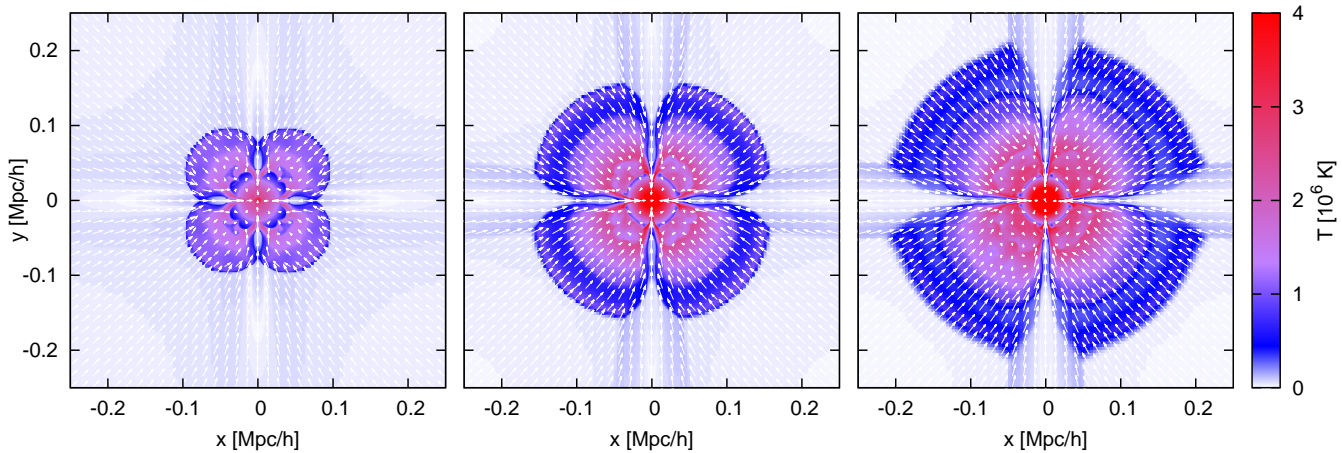}}
\caption{Time evolution of the temperature distribution of the halo. Shown are snapshots at expansion factors of a = (0.4,0.5,0.6) corresponding to redshifts of z = (1.5,1.0,0.6667). Again, the figure uses a linear scale. The vectors visualize the gas velocity in the plane of the slice. [\emph{See the electronic edition of the Journal for a color version of this figure.}]}
\label{fTimeSliceHalo}
\end{figure*}

In this section, we consider the formation and the evolution of the collapsed structure forming in the center of our simulation box which can be assigned to a gaseous halo. This halo forms at the intersection of the three filaments. Concerning the halo, our assumption about the identical distributions of gas and DM must be considered with caution. During its evolution, the gas halo could undergo a significant concentration in comparison to the DM distribution. In particular, this may happen due to the cooling processes. Shock heating, the introduction of an artificial pressure floor, and heating by the UV background acts against that concentration, however. On average, the halo gas enclosed by the shocks follows a nearly isothermal distribution. Thus, the two distributions (DM and gas) should not differ significantly from each other. Concordantly, the potential in our simulations corrected for the gravitational action of the DM is still a sufficient good approximation and there is no necessity to fully include the DM into our simulations.  

With these preconditions in mind, the halo region of our simulation box offers nevertheless an excellent testing ground to study the \emph{hydrodynamical} effects influencing the accretion over time (exclusive of more complicated baryon physics, like star formation, fragmentation, turbulent motions etc.). In particular, the supply of the inner halo regions with cold gas can be studied in this way. The formation processes for the halo exhibits some differences to the formation process for a filament being described  in Sect. \ref{sForm}. The density of the forming halo is much higher than inside the filament, and as a result, much higher temperatures are reached. This is caused the by strong heating due to adiabatic compression and by stronger shocks, caused by the deeper potential. 

Fig. \ref{fTimeSliceHalo} shows the time evolution of the temperature distribution in the halo region. Initially, a hot and dense clump forms by adiabatic compression at the intersection of the filaments. When the potential well of this clump is deep enough, a shock forms at the interface toward the void region. This shock is moving outwards with time and affects also the adjacent filament regions. The high pressure of the surrounding halo medium leads to a constriction of the inflowing filament gas. Therefore, the accretion onto the halo can be distinguished by two modes: 
\begin{itemize}
	\item The gas continuously infalling from the void region onto the halo is shock-heated to high temperatures $T \gtrsim 10^6$ K. 
	\item The colder gas, forming the isothermal core of the filament, can propagate deep into the halo supplying the halo with cold gas. 
\end{itemize}
The shock around the halo forms earlier than the shock confining the filaments. Only at later times, the subsequent gas accreting onto the latter produces shocks. Those shocks confine the cold isothermal cores of the filaments. 

As already mentioned, the described mechanism shows strong similarities to the cold stream scenario. We can now identify the cold streams with the isothermal cores in our simulated pancakes and filaments. The formation and evolution history obtained by our simulations leads immediately to an important conclusion: The cold streams have not to penetrate the shock around the halo, but are forming and existing at the very beginning of the evolution. They are attached to the halo even before any shock fronts around the halo have formed. The later forming shocks are influencing only the shape of the already existing streams, i.e. the cross section of the attached filaments and of their cold cores. This compression of the filament gas streaming into the 
halo was already shown in fully cosmological simulations, e.g. \citet{Dekel09}, \citet{Keres09b}.

\begin{figure}
\centering
\resizebox{\hsize}{!}{\includegraphics{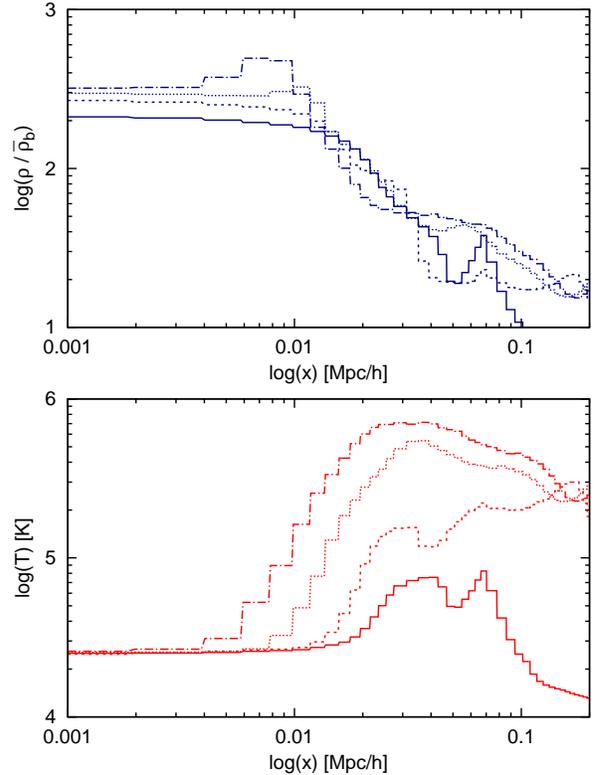}}
\caption{Temperature profiles through the cold stream in close vicinity to the halo for an initial perturbation amplitude of $A= 0.02$ and a perturbation scale of $L = 4$ Mpc/h. Shown are profiles at $z= 0$ including radiative cooling and heating by the UV background. Like in Fig. \ref{fProfileCool} (solid lines), we follow a line located entirely within the central sheet region. The profiles are displayed using logarithmic coordinate axes.
Displayed are profiles with decreasing distance to the center of the halo: 
340 kpc/h (\emph{solid line}).
240 kpc/h (\emph{dashed line}),
200 kpc/h (\emph{dottet line}),
and 180 kpc/h (\emph{dashed-dotted line}).
}
\label{fProfileStream}
\end{figure}

\subsection{Quenching by thermal conduction}

According to the above findings, the cross sections of the filament cores consisting of cold and dense gas are decreasing with increasing initial perturbation scale $L$.  The core size undergoes contraction along the filament's axis toward the gaseous halo and its gas content is time-dependent. In Figure \ref{fProfileStream} we show density and temperature profiles across the filament core (in $x$-direction, similar to the solid lines in Fig. \ref{fProfileCool}) for $L=4$ Mpc/h. The profiles are given at different locations near to the halo, where the shocked halo gas starts to influence the shape of the cold streams along the filament cores. It can be clearly noticed that with decreasing distance to the halo center the temperature gradient at the boundary of the isothermal core is significantly increasing and the cross section of the core is decreasing. High temperature gradients, however, might be accompanied by efficient thermal conduction. As discussed in Paper I, under certain conditions this may lead to a partial or entire evaporation of an initially cold region. An order-of-magnitude estimate gives for the case that thermal conduction may overcome the cooling, i.e. $-\nabla \cdot\mathbf{j} > \Lambda$, where $\mathbf{j}$ denotes the thermal flux and $\Lambda$ the cooling function:
\begin{equation}\label{eHeatEstimate}
	n_e \lambda_\mathrm{T} < 0.74 \times 10^{19} \,
	\left(\frac{T_\mathrm{c}}{10^4 \mathrm{K}}\right)^{-0.15}
	\left(\frac{T_\mathrm{shock}}{10^6 \mathrm{K}}\right)^\frac{7}{4}
	\mathrm{cm}^{-2} \;,
\end{equation}
where $n_e$ is the electron number density, $\lambda_\mathrm{T}$ is the characteristic scale for the temperature gradient, $T_\mathrm{c}$ is a temperature of the core of the stream, and $T_\mathrm{shock}$ is the temperature of the halo confining the stream. Under the prevailing thermal conditions and due to the UV background the gas is almost fully ionized $n_e \approx n_\mathrm{H}$. Saturation does not take place. The presence of a static magnetic field may considerably lower the efficiency of Spitzer's relation. However it is rather likely that the magnetic fields are not static but generated by turbulent chaotic motions. In that case, the heat conductivity can approach nearly Spitzer's coefficient \citep{Narayan01,Voigt04}. Thus Spitzer's relation for thermal conductivity is a good approximation. In addition we use quantities in order to estimate the lower limit of a possible effect. 

The contraction of the cold stream inside the halo is caused by the increased pressure, which is a result of the collapse process. At the outer boundary of the halo (given by the outwards moving shock) the spatial extend of the cold stream is still determined by the scaling relations obtained in Sect. \ref{sScaling}. The temperature of the shock of the halo, however, is caused by shock-heating alone. It can be therefore estimated from the transformation of the kinetic energy of the gas into thermal energy, when infalling onto the halo. This leads us to
\begin{equation}\label{eShockEstimate}
	\frac{k_\mathrm{B} T_\mathrm{shock}}{m_\mathrm{H}} \approx G \, \bar{\rho} \, L^2 
	\;\Rightarrow\;
	\left(\frac{T_\mathrm{shock}}{10^6 \mathrm{K}}\right)
	\approx 0.62 \, \left(\frac{L}{4 \mathrm{Mpc/h}}\right)^2 \;,
\end{equation}
where $m_\mathrm{H}$ denotes the mass of the hydrogen atom, $k_\mathrm{B}$ the Boltzmann constant, and $\bar{\rho}$ the background density. Thermal conduction becomes effective if the transition length $\lambda_\mathrm{T}$ will be of comparable size as the cold stream's cross section $\lambda_\circ$ or larger, i.e. $\lambda_\circ < \lambda_\mathrm{T}$. From Eq. (\ref{eHeatEstimate}) (transformed into dimensionless units) we obtain
\begin{equation}
	\left(\frac{n_\mathrm{Hc}}{1 \mathrm{cm}^{-3}} \right)^{-1}
	\left(\frac{T_\mathrm{c}}{10^4 \mathrm{K}}\right)^{-0.15}
	\left(\frac{T_\mathrm{shock}}{10^6 \mathrm{K}}\right)^\frac{7}{4}
	\left( \frac{\lambda_\circ} {1 \mathrm{Mpc/h}} \right)^{-1} \!\!\!
	> 2.3 \times 10^{-6}
\end{equation}
In Sect. \ref{sScaling} we obtained $n_\mathrm{Hc} \propto L^{5.7}$, $T_\mathrm{c} \propto L^{-1.2}$, and $\lambda_\circ \propto L^{-3.57}$. From Fig. \ref{fProfileStream} we obtain $\lambda_{\circ} \approx 0.025$ Mpc/h, $T_\mathrm{c} \approx 2.5 \times 10^4$, and $(\rho/\bar{\rho})_{\mathrm{c}} \approx 200$ (corresponding to $n_{\mathrm{Hc}} = 4.93 \times 10^{-5}$ cm$^{-3}$) at $L=4$ Mpc/h. With these values, as well as Eq. (\ref{eShockEstimate}), above inequality becomes:
\begin{equation}
	\left( \frac{L} {4 \mathrm{Mpc/h}} \right)^{-1.56} \lesssim 0.52
\end{equation}
When solving for $L$, we obtain a critical scale of 
\begin{equation}
	L_\mathrm{crit} \approx 6 \, \mathrm{Mpc / h}
\end{equation}
from which on thermal conduction may lead to an evaporation of the cold stream. This corresponds to a halo mass of $M_\mathrm{Halo} \approx 10^{13} M_\odot$.

The quantities $T_\mathrm{shock}$, $u_\mathrm{c,sheet}$, etc., which we considered in Sect. \ref{sScaling} depend on the redshift at which the shock formes \citep[see also][]{Sunyaev72}. Taking these dependencies into account we obtain for $\lambda_\circ \propto (1+z_\mathrm{s})^{-1}$ and for $\lambda_\mathrm{T} \propto (1+z_\mathrm{s})^{-5/4}$. Here $z_\mathrm{s}$ denotes the redshift of shock formation. If we include these dependencies into our estimate for the critical scale $L_\mathrm{crit}$, we obtain
\begin{equation}
	L_\mathrm{crit} \approx L_{\mathrm{crit},z=0} \, \left(1+z_\mathrm{s} \right)^{0.16} .
\end{equation}
This dependence on the shock formation time is rather moderate and makes the estimate for $L_\mathrm{crit}$ relatively robust.

Inside the halo, where the temperature gradients are even higher, thermal conduction becomes effective already at $L < L_\mathrm{crit}$, depending on the distance to the center of the halo. This should lead to a decrease of the penetration depth of the cold streams. Because of the drastic increase of computational complexity, we did not include thermal conduction into our simulations self-consistently. However, the above considerations together with the computational results in Paper I show, that under the existing conditions thermal conduction leads to an evaporation of the cold filament core. Concordantly, at initial perturbation lengths $L > L_\mathrm{crit}$, the formed cold streams are not longer able to penetrate deep into the halo, but evaporate if the halo forms a shock and starts to constrict the attached filament. Though the analytical considerations lead to a quite robust estimate for the critical $L_\mathrm{crit}$ the time-dependent evolution can only be obtained by future simulations.

\section{Summary and conclusions}\label{sConclusion}

In the present paper, we extend our investigations related to the formation of sheets in the intergalactic medium of the Universe as described in Paper I to a fully three-dimensional gas distribution. Again, we consider a highly symmetrical setup of initial and boundary conditions. We follow the evolution of one single perturbation mode sized according to the dimensions of the simulation box. In Paper I we have shown, that under conditions of shock heating the influence of almost all small-scale modes will be erased during the evolution of the largest mode provided the gas will be shocked on that scale. Taking into account only the largest perturbation mode, we are confronted with a problem similar to the cosmic variance problem in any cosmological simulation: The perturbation amplitudes related to the largest modes are sparsely sampled or even represented by a single value, only. Thus we chose an amplitude of the considered perturbation, which is not far from the cosmic power spectrum. We consider, however, different values for the initial amplitude, in order to show possible differences in the evolution of the perturbations. As in Paper I, we successively include different physical processes such as radiative cooling and heating due to the UV background into our simulations. In comparison, the outcome of our simulations, a three-dimensional configuration of sheets, filaments, and a gaseous halo, is much more complex. In particular, the higher degree of freedom leads to a complicated velocity field and well pronounced gas flows toward the maximum of gravitational potential. The main conclusions we derive from our three-dimensional simulations are as follows:

\begin{enumerate}

\item {\bf Formation and structure of filaments:} We are able to study the density and temperature distribution as well as the velocity field of the evolving filaments in great detail. Though the resolution of our simulation in the halo region is comparable with the obtained resolution in cosmological hydro simulations, the acquired resolution for the filaments (at low to intermediate gas densities) is in our simulations much higher. For these particular environment, our special setup, combined with the additional refinement according to the Jeans criterion, enables us to reach a resolution, which lies about three orders of magnitude in mass above the state-of-the-art IGM simulations (see also the footnote in the introduction). We get much more insight into the detailed structure, in particular the velocity field, of filaments of a scale up to 5 Mpc/h. In simulations with a lower spatial resolution these details would not be accessible. In particular, regions of high density and low temperature would be smoothed out by the pressure floor or a comparable mechanism to prevent a violation of the Truelove criterion. Less resolution makes necessary a higher pressure floor. Therefore those simulations become more and more similar to pure adiabatic simulations without cooling \citep[a similar conclusion was obtained in considerations by][]{Kang05}. For these reasons, the investigation given here might be considered as complementary to the existing hydrodynamical cosmological simulations. 
 
The gas streaming onto the filament forms a shock if the initial perturbation length exceeds $L = 4$ Mpc/h. The shock does not form everywhere at once, but develops first toward underdense regions, and successively covers the whole filament. In its center, the shock confined filament exhibits an isothermal core at temperatures of $T \approx 2 \times 10^4$ K. As for the one-dimensional case, sheets are forming, but they are not longer confined by shocks. The streaming motions toward the maximum of the gravitational potential cause a delayed collapse behavior and suppress shock-formation. 

The filament is provided with new gas essentially in two ways. On the one hand, gas accretes directly onto the filament (from the void region). This gas experiences a strong shock, which temperature is essentially given by the initial perturbation length. On the other hand, gas is flowing along the sheets towards the filaments. The shocks formed by this gas are closer to the filament's center and less pronounced. The thermodynamical state of the gas at the center of the filament is mainly determined by this gas.

The time of shock formation is governed by the value of the initial perturbation amplitude. A higher amplitude leads to an earlier shock.

\item {\bf The state and the evolution of the inner core of the filament:} One of the main characteristics of the filament is the state of the gas at its very center. Here, it forms a well confined core along the filament's axis consisting of cold and dense gas. The temperature of this gas is determined by the balance between heating due to the UV background and radiative cooling. In the early stage of the formation of the filament, due to the initial compression of the gas, radiative processes are very effective, i.e. cooling and heating times are short. Then, an equilibrium state between the (time-dependent) UV flux and cooling processes can be reached. During the further evolution, the time evolution of the UV flux and the overall density dilution due to the cosmic expansion moves this equilibrium state along the phase trajectory $T_\mathrm{c}(n_\mathrm{Hc})$. The quantities related to that core ($n_\mathrm{Hc}, T_\mathrm{c}, \lambda_\mathrm{c}$, etc.) exhibit a very strong dependence on the initial perturbation length $L$. For the filaments in the three-dimensional case this dependence is much stronger and the physical mechanisms are different compared to the one-dimensional case. For the gas accreting directly (\emph{not} via the sheets) the temperature scales again $\propto L^2$. This is a general outcome, and can be derived from the hydrodynamical equations as long as dissipative processes can be neglected \citep[see also][]{Sunyaev72}. 

It can be noticed from the formation process of the inner region of a filament that the state of the core is determined by the gas inflowing along the sheets. The dominant gas pressure is in this case not longer determined by the gas temperature in the shock surrounding the filament, but by the dynamical pressure of the gas streaming in via the sheets. Taking that into account, we are able to derive analytically the scaling relations confirming the results obtained from our simulations. The initial amplitude does not enter these relations. It determines solely the moment of time when the shocks and the cold core form. However, since the UV flux $j_0$ is time dependent and defines the equilibrium state at any moment, the moment when the shock forms also defines the starting point of the equilibrium track along $T_\mathrm{c}(n_\mathrm{Hc}, j_0)$. The obtained scaling relations allow for a reliable extrapolation of the results beyond $L = 6$ Mpc/h where the obtainable resolution becomes unsufficient. In particular, the strong $L$-dependence of the core size $\lambda_\mathrm{c}$ leads to an extremely constricted core for large $L > 14$ Mpc/h at redshifts $z \approx 0$. 

\item {\bf The cold streams:} The velocity of the gas, which belongs to the cold core is aligned along the local filament axis and is directed toward the center of the collapsing halo. Concordantly, these streams provide the inner halo region with cold and dense gas. The overall properties of these streams correspond to the cold streams observed in hydrodynamical simulations of galaxy formation \citep[see e.g.][]{Nagai03}. Hence, if one attributes the halo to a galaxy, this gas would significantly impact on the star formation history of this galaxy. We find these streams to be an essential characteristic of our configuration, and they exist from the very beginning of the collapse process. Being a part of the core of the filament, the stream follows the same scaling relations. Therefore, the stream becomes narrower and denser with increasing $L$. The influence of the outward moving accretion shock of the halo results in a further constriction of the stream. This effect becomes stronger with decreasing distance to the halo center.

The total amount of gas which can be accreted via cold streams, is determined by the same processes which control the core of the filaments. The main constituent in this context, is the interplay between radiative cooling and heating by the UV background. The combination of these processes with the particular hydrodynamic conditions, given by the characteristic scales of the system, result in the scaling relations discussed above. Besides the radiative cooling and the heating by the UV background, additional processes acting on the energetic budget of the gas, e.g. radiative feedback from the forming galaxy, may have an impact on the spatial extend of the stream as well as on the amount of transported gas. 

After the filament has formed, the amount of mass inside the filaments core is decreasing with time. The gas infalling from the void region onto the filament is shocked and can therefore not reach the core region, anymore. The cooling time for the gas at high temperature and low density has become too large. At some stage of evolution, also the gas streaming along the sheets does not reach the filament's core region. This happens when the dynamic pressure of the gas flow along the sheets becomes equal or less the pressure of the cold and dense gas of the filament's core. Therefore, the core is shielded from further gas supply. In the result, the total reservoir of cold gas available for star formation processes form cold stream accretion is limited to a well defined fraction. This gas fraction is determined soon after the formation of the filaments. With time, the core gas is draining off toward the gaseous halo and the mass content of the filament cores is expected to decrease continuously. This is a consequence of structure formation only.

Taking into account the full initial perturbation spectrum for the DM which was not considered here might alter some results. In particular, the gas content in the cold streams available for gas supply of the halo may considerably decrease due to the formation of collapsed objects within the filament's core region. As in all cosmological simulations, the large-scale power is cut due to the fixed size of the simulation box. This may have impact on the time evolution. Taking into account the action of the large-scale modes of the perturbation spectrum beyond the actual box size, tidal fields, angular momentum transfer etc. will modify the velocity field and in particular the geometrical shape of the cold streams. Therefore, the gas flow into the halo will also change with time.

\item {\bf Termination of cold streams:}
The enhanced pressure of the shocked gas inside the halo leads to a subsequent constriction of the cold streams reaching the inner halo region. This effect increases with decreasing distance to the hot and dense center of the halo and is caused by the adiabatic contraction of the gas. It needs the additional influence of thermal conduction to prevent the cold gas from streaming till almost the center of the gas halo. At the outer vicinity of the halo, large temperature gradients occur. These gradients become even stronger towards the center of the halo. As a result, thermal conduction is effective, and the heating by the thermal flux toward the cold stream region is able to exceed the radiative cooling there. Using the scaling relations for the core of the filament, we obtain that for initial perturbation lengths $L > 6$ Mpc/h, which corresponds to a halo mass of about $M_\mathrm{halo} \approx 10^{13} M_\odot$, the cold stream evaporates at the boundary of the accretion shock and is therefore not able to exist inside the halo, anymore. For intermediate perturbation lengths $2$ Mpc/h $< L < 6$ Mpc/h a partly evaporation of the streams, caused by  constriction of the stream inside the halo, may occur. This process should start near the central region, first. Thus, a termination of cold gas supply may happen for smaller scales as well. The corresponding mass range for which the cold streams can reach the inner halo is $3 \times 10^{11} M_\odot < M_\mathrm{halo} < 10^{13} M_\odot$.

These results are calculated for $z \approx 0$. If the initial perturbation amplitude is higher, the shock formation and the collapse of the halo start considerably earlier, i.e. at higher redshifts. As a consequence, the physical density in the filament's core is higher. The characteristic length for thermal conduction $\lambda_\mathrm{T}$ decreases with increasing redshift stronger than the core size $\lambda_\circ$. In the result, the estimate for the critical length scale $L_\mathrm{crit}$ gives moderately higher values $L_\mathrm{crit} \propto (1+z_\mathrm{s})^{0.1}$, where $z_\mathrm{s}$ denotes the time of shock formation. This means that if shocks happen at higher redshifts, also the upper bound for the mass of the halos which can still be provided with dense cold gas via the attached filaments remains almost unchanged. We conclude, that for halo masses exceeding the above given range $3 \times 10^{11} M_\odot < M_\mathrm{halo} < 10^{13} M_\odot (1+z_\mathrm{s})^{0.3}$ a supply of the inner halo with cold gas will be terminated. 

\item {\bf Observations:}  The temperature of the shocked gas in the filaments scales as $\propto L^2$. Thus, for perturbation scales of  $L \approx 10$ Mpc/h the temperature of the filament gas reaches almost $10^6$ K. Those filaments can be attributed to the bulk of the WHIM as predicted by hydrodynamical cosmological simulations. The observation of WHIM filaments at these scales remains challenging. The core region of the filament becomes too small and the probability that a line of sight intersects with this particular region becomes extremely small, as well. Thus, the filaments are hot almost everywhere and do not exhibit significant variations which could serve as a characteristic tracer for observations. However, if by chance the line of sight hits the core region then absorption by hydrogen could be observed at high column densities. Since the spatial extension of the core scales as $\propto L^{-3.57}$ and the density as $\propto L^{5.7}$, the column density of neutral hydrogen scales as $\propto L^{7.83}$ as long the core is still optically thin.Using the values for $L=4$ Mpc we obtain an estimate for the column density $N_{\ion{H}{I}} = n_{\ion{H}{I}} \, \lambda_\circ \approx 2 \times 10^{14} \, (L/4\, \mathrm{Mpc})^{7.83} \, (1+z_\mathrm{s})^5$. For $z \leq 3$ and for $L \leq L_\mathrm{crit}$ the assumption of an optical thin core holds. At higher redshifts or for considerably larger $L$ the core may become shielded against the UV radiation. In this case, radiative cooling is not longer balanced by UV heating and further contraction will occur. However, at some stage the medium becomes optically thick with respect to recombination radiation. The radiation is not longer able to leave the cold core freely, but interacts significantly with the gas before leaving the core region. Thus, for a primordial matter composition, the cooling becomes less efficient \citep[see][]{Gnat07} and for a full description radiative transfer effects must be taken into account. In this context, and as a a goal for future studies, it is needed to consider possible ionization states and the corresponding column densities for heavy elements. Temperature and density profiles with partly steep gradients may cause characteristic absorption features. The main uncertainty, however, remains the abundance of the heavy elements in the filaments far from halos/galaxies. 
\end{enumerate}

\section*{Acknowledgments}
We thank the anonymous referee for his numerous valuable remarks and suggestions. JSK acknowledges A. Partl, T. Doumler, S. Knollmann and J.-C. Mu\~nos-Cuartas for help and useful discussions. The simulations were performed using a modified version of \texttt{RAMSES}. We want do acknowledge R. Teyssier for developing and maintaining the code and thank him for making it publicly availlable to the community. JSK was supported by the Deutsche Forschungsgemeinschaft under the project MU 1020/6-4.s and by the German Ministry for Education and Research (BMBF) under grant FKZ 05 AC7BAA.

\bibliographystyle{mn2e}
\bibliography{filament}{}	

\bsp

\label{lastpage}

\end{document}